\documentclass{PoS}

\usepackage{latexsym}
\usepackage[italian,english]{babel}
\usepackage[latin1]{inputenc}
\usepackage{epstopdf}
\usepackage[fleqn]{amsmath}
\usepackage{amsfonts}
\usepackage{amssymb}
\usepackage{amsmath}
\usepackage{amsthm}
\usepackage{bm}
\usepackage{amsopn}
\usepackage{xspace}
\usepackage{array}
\usepackage{epsfig}
\usepackage{rotating}
\usepackage{subfig}
\usepackage{url}
\usepackage{graphicx}
\usepackage[T1]{fontenc}
\usepackage{lmodern}
\usepackage[numbers]{natbib}
\usepackage{abbr} 
\usepackage{longtable}
\usepackage{textcomp}
\usepackage{pdflscape}
\usepackage{rotating}
\usepackage{setspace}
\usepackage{footmisc}
\usepackage{booktabs}
\usepackage{multirow}
\usepackage{booktabs}
\usepackage[section]{placeins}
\title{On the connection between radio mini-halos and gas heating in cool-core clusters}

\ShortTitle{Radio mini-halos and gas heating in cool-core clusters}

\author{\speaker{Luca Bravi}\\
        INAF-ORA, via Gobetti 101, I-40129 Bologna, Italy\\
        E-mail: \email{luca.bravi@studio.unibo.it}}

\author{{Myriam Gitti}\\
		Dipartimento di Fisica e Astronomia, Universit\'a di Bologna, via Ranzani 1, I-40127 Bologna, Italy\\
        INAF-ORA, via Gobetti 101, I-40129 Bologna, Italy}

\author{{Gianfranco Brunetti}\\
	\smallskip
		INAF-ORA, via Gobetti 101, I-40129 Bologna, Italy}

\abstract{In this work, we present a study of the central regions of cool-core clusters hosting radio mini-halos, which are diffuse synchrotron sources extended on cluster-scales surrounding the radio-loud brightest galaxy. We aim to investigate the interplay between the thermal and non-thermal components in the intracluster medium in order to get more insights into these radio sources, whose nature is still unclear. It has recently been proposed that turbulence plays a role for heating the gas in cool cores. A correlation between the radio luminosity of mini-halos, $\nu P_{\nu}$, and the cooling flow power, $P_{\rm CF}$, is expected in the case that this turbulence also plays a role for the acceleration of the relativistic particles. We carried out a homogeneous re-analysis of X-ray \textit{Chandra} data of the largest sample of cool-core clusters hosting radio mini-halos currently available ($\sim20$ objects), finding a quasi-linear correlation, $\nu P_{\nu} \propto P_{\rm CF}^{0.8}$. We show that the scenario of a common origin of radio mini-halos and gas heating in cool-core clusters is energetically viable, provided that mini-halos trace regions where the magnetic field strength is $B \gg 0.5 \mu G$.}

\FullConference{EXTRA-RADSUR2015 (*)\\
	20--23 October 2015\\
	Bologna, Italy
	
	\bigskip
	\hrule
	\bigskip
	
	\textnormal{(*) This conference has been organized
		with the support of the Ministry of Foreign Affairs
		and International Cooperation, Directorate General
		for the Country Promotion (Bilateral Grant Agreement
		ZA14GR02 - Mapping the Universe on the Pathway to
		SKA)}
}

\begin{document}
\section{Introduction}
\vspace{-0.1in}
\noindent
The central intracluster medium (ICM) of cool-core (CC) clusters must experience some kind of heating to balance cooling  (\cite{Peterson_2006}). The most promising source of heating has been identified as feedback from active galactic nucleus (AGN) (\citep[e.g.,][]{Gitti_2012}). At the same time, mechanically powerful AGN are likely to drive turbulence in the central ICM which may contribute to gas heating. For example, \cite{Zhuravleva_2014} found that AGN-driven turbulence must eventually dissipate into heat and it is sufficient to offset radiative cooling. On the other hand, such turbulence can also play a role in the origin of the diffuse synchrotron emission observed in a number of CC clusters in the form of radio mini-halos (MH), whose origin is still unclear (e.g, \cite{Brunetti_2014}). A possibility is that they form through the re-acceleration of relativistic particles by turbulence (\citep[][]{Gitti_2002,Zuhone_2013}). Alternatively, they may be of hadronic origin (\citep{Pfrommer_2004}).

In this work we present the result of a homogeneous re-analysis of \textit{Chandra} data of the largest collection of radio MHs currently known, aimed at investigating a possible re-acceleration scenario where the turbulence is responsible for both the origin of MHs and for quenching cooling flows. We overcome the limitations of the previous studies by exploiting the increased statistics of known radio MHs. 

\vspace{-0.1in}
\section{Mini-halo sample selection and X-ray data analysis}
\vspace{-0.1in}
\noindent
Our sample is obtained from the list of 21 mini-halos reported in \cite{Giacintucci_2014}, who recently selected a large collection of X-ray-luminous clusters from the \textit{Chandra} ACCEPT\footnote{Archive of \textit{Chandra} Cluster Entropy Profiles Table} sample (\cite{Cavagnolo_2009}) and discovered four new MHs. We further included the new MH detection in the Phoenix cluster (\cite{vanWeeren_2014}). In order to determine the physical properties of the thermal ICM in the region \hfill where \hfill the \hfill diffuse \hfill radio \hfill emission \hfill is \hfill present, \hfill we \hfill extracted \hfill a \hfill spectrum \hfill inside \hfill the \par
\begin{table*}[hb!]\tiny
	\caption{Properties of our sample of mini-halo clusters.}
	\label{tab:radio_MH}      
	{\centering
		\begin{tabular}{p{2.5cm}p{1cm}ccccc}
			\midrule
			Cluster name & \textit{z} & $S_{{\rm MH \, [1.4GHz]}}$ & $R_{{\rm MH}}$ & $\nu P_{\nu \, {\rm [1.4GHz]}}$ & $P_{\rm CF}$ & $L_{{\rm NT}, B =1\, \mu\mathrm{G}}$\\
			& & (mJy) & (kpc) & $\mathrm{10^{40} \; erg \, s^{-1}}$ & $10^{44} \, \mathrm{erg \, s^{-1}}$ & $\mathrm{10^{40} \; erg \, s^{-1}}$\\
			\midrule
			2A 0335+096 & 0.035 & $21.1\pm 2.1$ & 70 & $0.08\pm 0.01$ & $ 0.32\pm 0.02$ & $10.8\pm 1.1$\\
			A 2626 & 0.055 & $18.0\pm 1.8$ & 30 & $0.19\pm 0.01$ & $0.02\pm 0.01$ & $27.6\pm 2.0$\\
			A 1795 & 0.063 & $85.0\pm 4.9$ & 100 & $1.11\pm 0.07$ & $0.10\pm 0.04$ & $159.9\pm 10.1$\\
			ZwCl 1742.1+3306 & 0.076 & $13.8\pm 0.8$ & 40 & $0.28\pm 0.01$ & $0.09\pm 0.02$ & $42.4\pm 2.1$\\
			A 2029 & 0.077 & $19.5\pm 2.5$ & 270 & $0.39\pm 0.06$ & $0.07\pm 0.05$ & $59.6\pm 8.5$\\
			A 478 & 0.088 & $16.6\pm 3.0$ & 160 & $0.45\pm0.08$ & $ < 0.34$ & $70.8\pm 13.3$\\
			A 2204 & 0.152 & $8.6\pm 0.9$ & 50 & $0.76\pm 0.07$ & $< 0.06$ & $148.3\pm 13.7$\\
			RX J1720.1+2638 & 0.159 & $72.0\pm 4.4$ & 140 & $7.46\pm 0.45$ & $< 0.20$ & $1497\pm 90$\\
			RXC J1504.1-0248 & 0.215 & $20.0\pm 1.0$ & 140 & $3.78\pm 0.20$ & $3.55\pm 2.46$ & $908.7\pm 47.1$\\
			A 2390 & 0.228 & $28.3\pm 4.3$ & 250 & $6.24\pm 0.94$ & $< 1.34$ & $1562\pm 235$\\
			A 1835 & 0.252 & $6.1\pm 1.3$ & 240 & $1.66\pm 0.35$ & $2.25\pm 0.90$ & $449.0\pm 94.3$\\
			MS 1455.0+2232 & 0.258 & $8.5\pm 1.1$ & 120 & $2.45\pm 0.32$ & $1.11\pm 0.91$ & $672.2\pm 88.3$\\
			ZwCl 3146 & 0.290 & $\sim 5.2$ & 90 & $1.95\pm0.01$ & $1.58\pm 0.90$ & $588.7\pm 0.04$\\
			RX J1532.9+3021 & 0.345 & $7.5\pm 0.4$ & 100 & $4.69\pm 0.24$ & $3.41\pm 1.98$ & $1668\pm 84.7$\\
			MACS J1931.8-2634 & 0.352 & $47.9\pm 2.8$ & 100 & $28.0\pm 1.7$ & $7.17\pm 3.30$ & $10162\pm 610$\\
			RBS 797 & 0.354 & $5.2\pm 0.6$ & 120 & $3.08\pm 0.34$ & $< 3.00$ & $1125\pm 123$\\
			MACS J0159.8-0849 & 0.405 & $2.4\pm 0.2$ & 90 & $1.95\pm 0.20$ & $< 3.90$ & $826.3\pm 82.6$\\
			MACS J0329.6-0211 & 0.450 & $3.8\pm 0.4$ & 70 & $3.98\pm 0.42$ & $< 3.36$ & $1896\pm 200$\\
			RX J1347.5-1145 & 0.451 & $34.1\pm 2.3$ & 320 & $35.8\pm 2.5$ & $61.3\pm 29.1$ & $17132\pm 1198$\\
			Phoenix & 0.596 & $6.8\pm 2.0$ & 176 & $14.1\pm 4.2$ & $< 121.81$ & $9832\pm 2920$\\
			\midrule
			\multicolumn{7}{l}{%
				\begin{minipage}{0.98\textwidth}
					\tiny \textbf{Notes:} Col. (1): Cluster name. Col. (2):
					Redshift. Col. (3): Mini-halo flux density at 1.4 GHz from \cite{Giacintucci_2014}, except in the case of Phoenix where the value was estimated from the observations at 610 MHz (\cite{vanWeeren_2014}) by assuming a spectral index of $\alpha = 1.1$. Col. (4): Average radius of the mini-halo. Col. (5): Radio power of mini-halos at 1.4 GHz (in terms of integrated radio luminosity, $\nu P_{\nu}$). Col. (6): Cooling flow power estimated as $P_{{\rm CF}} = \frac{\dot{M}kT}{\mu m_{p}}$ inside $R_{\rm MH}$. Col (7): Non-thermal luminosity estimated for a reference value of the magnetic field strength of $B =1\, \mu$G.
				\end{minipage}}
				\\
			\end{tabular}}
		\end{table*}
\newpage
\noindent radius of the MH ($R_{\rm MH}$) for each cluster of the sample. We derived the temperature, \textit{kT}, and the mass deposition rate, $\dot{M}$, from detailed spectral analysis (\cite{Bravi_2016}). The mini-halo sample, the value of the physical parameters with their 90\% confidence level derived for each cluster are listed in Table~\ref{tab:radio_MH}.
\vspace{-0.1in}
\section{The correlation $\nu P_{\nu}$ - $P_{\rm CF}$}
\vspace{-0.1in}
\noindent
Gitti et al. (\cite{Gitti_2004,Gitti_2012}) found a correlation between the radio power of mini-halos at 1.4 GHz (in terms of integrated radio luminosity, $\nu P_{\nu}$), and the cooling flow power, $P_{{\rm CF}}$. The maximum power $P_{\rm CF}$ available in the cooling flow can be estimated assuming a standard cooling flow model and it corresponds to the $p dV/dt$ work done on the gas per unit time as it enters the cooling radius, defined as the radius at which the cooling time is equal to the age of system ($P_{\rm CF} =  \frac{\dot{M} kT}{\mu m_{{\rm p}}}$).
In order to compare powers emitted inside the same volume, i.e. that of MH, in this work we estimated $P_{{\rm CF}}$ inside $R_{\rm MH}$.\\
\begin{figure}[!h]
	\centering
	\includegraphics[width=7.5cm]{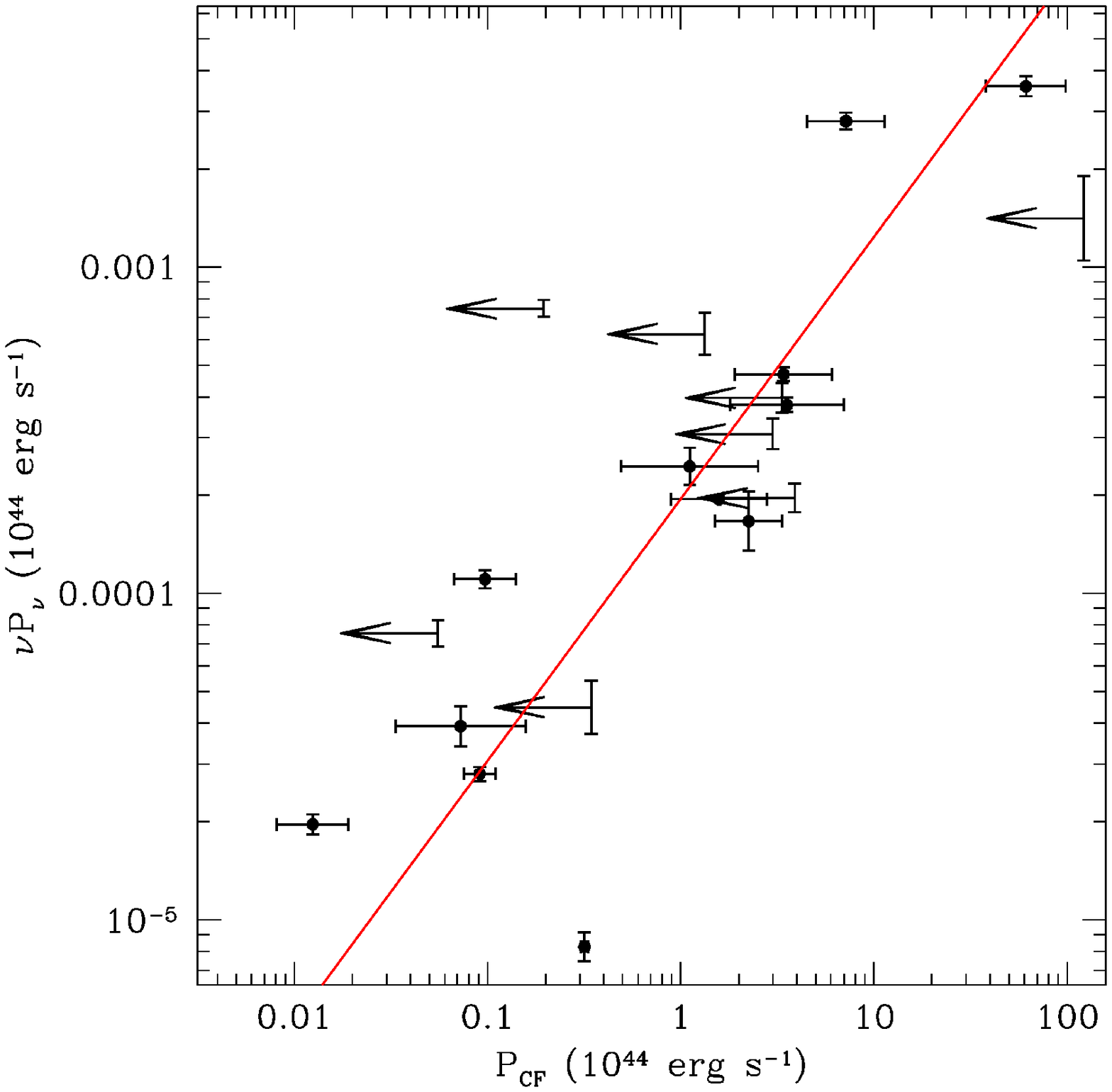}
	\includegraphics[width=7.5cm]{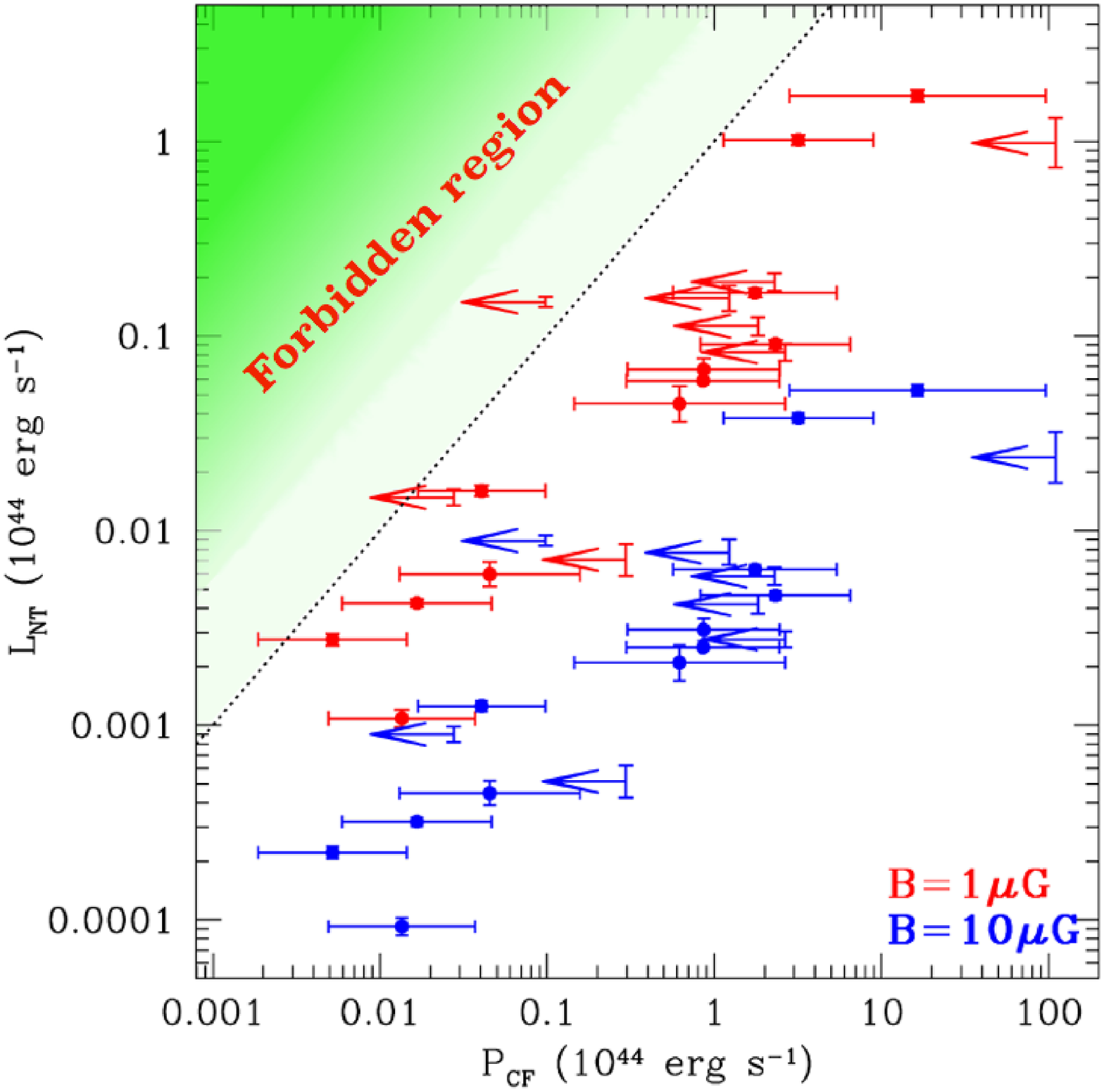}
	\caption{\footnotesize{\textit{Left panel:} correlation between the radio emitted power of mini-halos at
			1.4 GHz, in terms of $\nu P_{\nu}$, and the cooling flow power,
			$P_{{\rm CF}}$, for our sample of mini-halo clusters (see Table~\ref{tab:radio_MH}). The black arrows show the upper limits when $P_{{\rm CF}}$ is not
			constrained. The red line represents the best-fit relation determined without the upper limits on $P_{{\rm
					CF}}$. \textit{Right panel:} correlation between $L_{\rm NT}$ and $P_{\rm CF}$ for our
			MH cluster sample. The red and blue points were calculated assuming $B =1\, \mu$G and $B =10\, \mu$G, respectively. The black dotted line represents the 1:1 relation. The upper side of 1:1 relation is the forbidden region in our turbulent re-acceleration scenario, where $L_{\rm NT} > P_{\rm CF}$.}}\label{fig:pcf}
\end{figure}By using the 12 clusters of our sample for which the value of $P_{{\rm CF}}$ is constrained, we found a quasi-linear correlation in the form:
\begin{equation}
\log(\nu P_{\nu})=[(0.80\pm 0.13) \cdot \, \log(P_{\rm CF})]-(3.70\pm 0.11)
\label{eq:fit1}
\end{equation}
The correlation is shown in Fig.~\ref{fig:pcf} (\textit{left panel}). This suggests a connection between the energy reservoir in cooling flows and that associated to the non-thermal components powering radio MHs, i.e. relativistic particles and magnetic field, confirming the previous results obtained by \cite{Gitti_2004,Gitti_2012}.

\vspace{-0.1in}
\section{Discussion and conclusion}
\vspace{-0.1in}
\noindent
The heating mechanisms proposed to solve the so-called ``cooling flow problem'' envision a gentle energy dissipation which is distributed in the core (comparable to the size of MHs), with a heating rate that cannot be much greater than the cooling power, otherwise cool cores would be disrupted. In particular, it has recently been proposed that turbulent dissipation may compensate gas cooling losses thus keeping cluster cores in an approximate steady state (\cite{Zhuravleva_2014}). Turbulence is also proposed as an important player for the origin of MHs (leptonic models, \cite{Gitti_2002,Zuhone_2013}). In this work we argue that particle acceleration powering the non-thermal emission from MHs and gas heating in CCs are due to the dissipation of the same turbulence. In this case, $P_{{\rm CF}}$ provides an upper limit to the non-thermal (synchrotron and inverse Compton) luminosity $L_{{\rm NT}}$ (see Equation (5) in \cite{Bravi_2016}) generated in the MH region. $L_{{\rm NT}}$ depends on the magnetic field intensity in the MH region that sets the fraction of $L_{{\rm NT}}$ that goes into synchrotron radiation. Fig.~\ref{fig:pcf} (\textit{right panel}) shows the $L_{{\rm NT}}$ of MHs of our sample versus $P_{{\rm CF}}$ for two different values of the magnetic field \textit{B}. In the case of $B =10\, \mu$G the non-thermal luminosity of the MHs remain distant from the forbidden region and the proposed turbulent re-acceleration scenario is energetically consistent. Instead, for $B < 0.5\, \mu$G, we find that $L_{\rm NT} \gtrsim P_{\rm CF}$ and the scenario becomes not plausible. Therefore the limit $P_{{\rm CF}} \gg L_{\rm NT}$ allows us to set a corresponding lower limit $B > 0.5\, \mu$G to the typical magnetic field in MHs.

To summarize, we have overcome the limitations in the previous studies by exploiting the increased statistics of known radio MHs that allows us to obtain further insights on their origin. Future observations with ASTRO-H in the hard X-ray and Faraday rotation studies with the new radio facilities are fundamental to increase the number of known MHs and achieve a full understanding of the mechanisms for their origin (see e.g. \cite{Gitti_2015} for a discussion about the possibility offered by future SKA radio surveys).

\vspace{-0.1in}
\footnotesize{
\bibliographystyle{apj_short_etal}
\bibliography{bibliography}}
%

\end{document}